\documentclass[a4]{article}
\usepackage[margin=2cm]{geometry}

\usepackage{color}
\usepackage{amssymb,amsmath}
\usepackage{graphicx}
\usepackage{subfigure}
\usepackage{wrapfig}
\usepackage{epsfig}
\usepackage{endnotes}
\usepackage[normalem]{ulem}
\usepackage{url}
\usepackage{hyperref}
\usepackage{color}

\begin{document}

\date{January 8, 2015}

\title{Collective Mind, Part II: Towards Performance- \\ and Cost-Aware Software Engineering \\ as a Natural Science}

\author{Grigori Fursin$^{1}$, Abdul Memon$^{2}$, Christophe Guillon$^3$, and Anton Lokhmotov$^4$ \\
\\
$^1$~cTuning foundation, France ; $^2$~UVSQ, France  \\
$^3$~STMicroelectronics, France ; $^4$~ARM, UK ; \\
}

\maketitle

{\it Presented at the 18th International Workshop on Compilers for Parallel Computing (CPC'15), London, UK}

\begin{abstract}
Nowadays, engineers have to develop software often without even
knowing which hardware it will eventually run on in numerous
mobile phones, tablets, desktops, laptops, data centers,
supercomputers and cloud services.
Unfortunately, optimizing compilers are not keeping pace with
ever increasing complexity of ever changing computer systems
anymore and may produce severely underperforming executable codes
while wasting expensive resources and energy.

We present the first to our knowledge practical, collaborative
and publicly available solution to this problem.
We help the software engineering community gradually implement 
and share light-weight wrappers around any software piece with 
more than one implementation or optimization choice available.
These wrappers are connected with a public
\href{http://c-mind.org}{Collective Mind autotuning
infrastructure and repository of knowledge} to continuously
monitor all important characteristics of these pieces
(\emph{computational species}) across numerous existing hardware
configurations in realistic environments together with randomly
selected optimizations.
At the same time, \href{http://c-mind.org/cmn}{Collective Mind Node}) 
allows to easily crowdsource time-consuming autotuning across
existing Android-based mobile device including commodity
mobile phones and tables. 

Similar to natural sciences, we can now continuously track all
winning solutions (optimizations for a given hardware such
as compiler flags, OpenCL/CUDA/OpenMP/MPI/skeleton parameters,
number of threads and any other exposed by users) that minimize
all costs of a computation (execution time, energy spent, code
size, failures, memory and storage footprint, optimization time,
faults, contentions, inaccuracy and so on) of a given species
on a Pareto frontier along with any unexpected behavior
at \href{http://c-mind.org/repo}{c-mind.org/repo}
\footnotetext[1]{We are now developing a smaller, simpler and
faster version of our collaborative infrastructure and repository
(Collective Knowledge or CK). Preview is available
at~\url{http://github.com/ctuning/ck} and
\url{http://cknowledge.org/repo}.}.
Furthermore, the community can continuously classify solutions, 
prune redundant ones, and correlate them with various features
of software, its inputs (data sets) and used hardware either
manually (similar to Wikipedia) or using available big data 
analytics and machine learning techniques.

Our approach can also help computer engineering community create 
the first public, realistic, large, diverse, distributed, representative,
and continuously evolving benchmark with related optimization
knowledge while gradually covering all possible software and
hardware to be able to predict best optimizations and improve
compilers depending on usage scenarios and requirements.
Such continuously growing collective knowledge accessible via
simple web service can become an integral part of the practical
software and hardware co-design of self-tuning computer systems 
as we demonstrate in several real usage scenarios validated 
in industry.\footnote[2]{{\it Corresponding author's email: Grigori.Fursin@cTuning.org}}.

\end{abstract}


\section{Introduction and Related Work}
\label{introduction}

There is an impressive and ever increasing number of diverse
computer systems available nowadays on the market.
They all vary in performance, size, power consumption,
reliability, price and other characteristics depending
on numerous available hardware features such as processor
architecture, number of cores, availability of specialized
hardware accelerators, working frequency, memory hierarchy and
available storage.
As a result, engineers often have to develop software that may
end up running across different, heterogeneous and possibly
virtualized hardware in multiple mobile devices, desktops, 
HPC servers, data centers and cloud services.
Such a rising complexity of computer systems and limited
development time usually force software engineers to rely almost
exclusively on existing compilers, operating systems and run-time
libraries in a hope to deliver the fastest, smallest, most power
efficient, scalable and reliable executable code across all
available hardware.
Unfortunately, this complexity of ever changing hardware also
made development of compilers very challenging.
They nowadays include hundreds of optimizations and often
fail to produce efficient code while wasting expensive resources
and energy~\cite{atlas, europar97x, citeulike:1671417,
Hall:2009:CRN:1461928.1461946, fursin:hal-01054763}.

Numerous autotuning, run-time adaptation, genetic and machine
learning techniques (including our own) have been introduced
in the past two decades to help software engineers optimize their
applications for rapidly evolving hardware~\cite{atlas,
europar97x, CGJ1997, Nis1998, fftw, CSS99, VE00, KKO2000, FOK02,
SAMP2003, Tapus:2002:AHT:762761.762771, vista, spiral, LCYP04,
la2004, FOTP2005, PE2006, HE2008, BCCP2008, JGVP2009,
Ansel:2009:PLC:1542476.1542481, Mars:2010:CAE:1772954.1772991,
DBLP:conf/cc/MooreC13, DBLP:conf/cf/ShenVSAS13,
Miceli:2012:APA:2451764.2451792,
Manotas:2014:SSE:2568225.2568297}.
These techniques usually demonstrate that it is possible
to improve various characteristics of existing software
by automatically and empirically searching and predicting 
better combinations of optimizations.
Such optimizations typically include compiler flags,
optimization parameters and their orders, different algorithm
implementations, run-time scheduling policies, working frequency
among many other choices.
The resulting 10\% to 10x performance improvements of frequently
executed programs can already reduce usage costs of large data
centers and supercomputer operators by thousands of dollars per
year.
These improvements can also considerably improve overall 
performance, battery life and storage space in mobile devices.

Nevertheless, in spite of so many promising results and research
advances in the past twenty years, we still witness the rising
number of reports, complaints, public discussions and research
articles showing that existing compilers are missing many optimization
opportunities for popular software~\cite{llvm-dev-mailing-list, 
gcc-dev-mailing-list, phoronix-forum}.
This is caused by a fundamental and yet unsolved problem -
design and optimization spaces are already too large and
continue growing.
Therefore, exhaustive exploration of the whole space 
to find the best solution is simply impossible.
Practically all recent long-term research visions acknowledge
above problem~\cite{xciteulike:1671417,
Dongarra:2011:IES:1943326.1943339, prace,
Hall:2009:CRN:1461928.1461946, hipeac_roadmap}.
They even advocate that some radically new approaches should
be invented to be able to continue building faster, more power
efficient and reliable software and hardware by 2020 while
somehow decreasing all development, optimization and usage costs.

\begin{figure}[htb]
  \centering
   \includegraphics[width=4.5in]
   {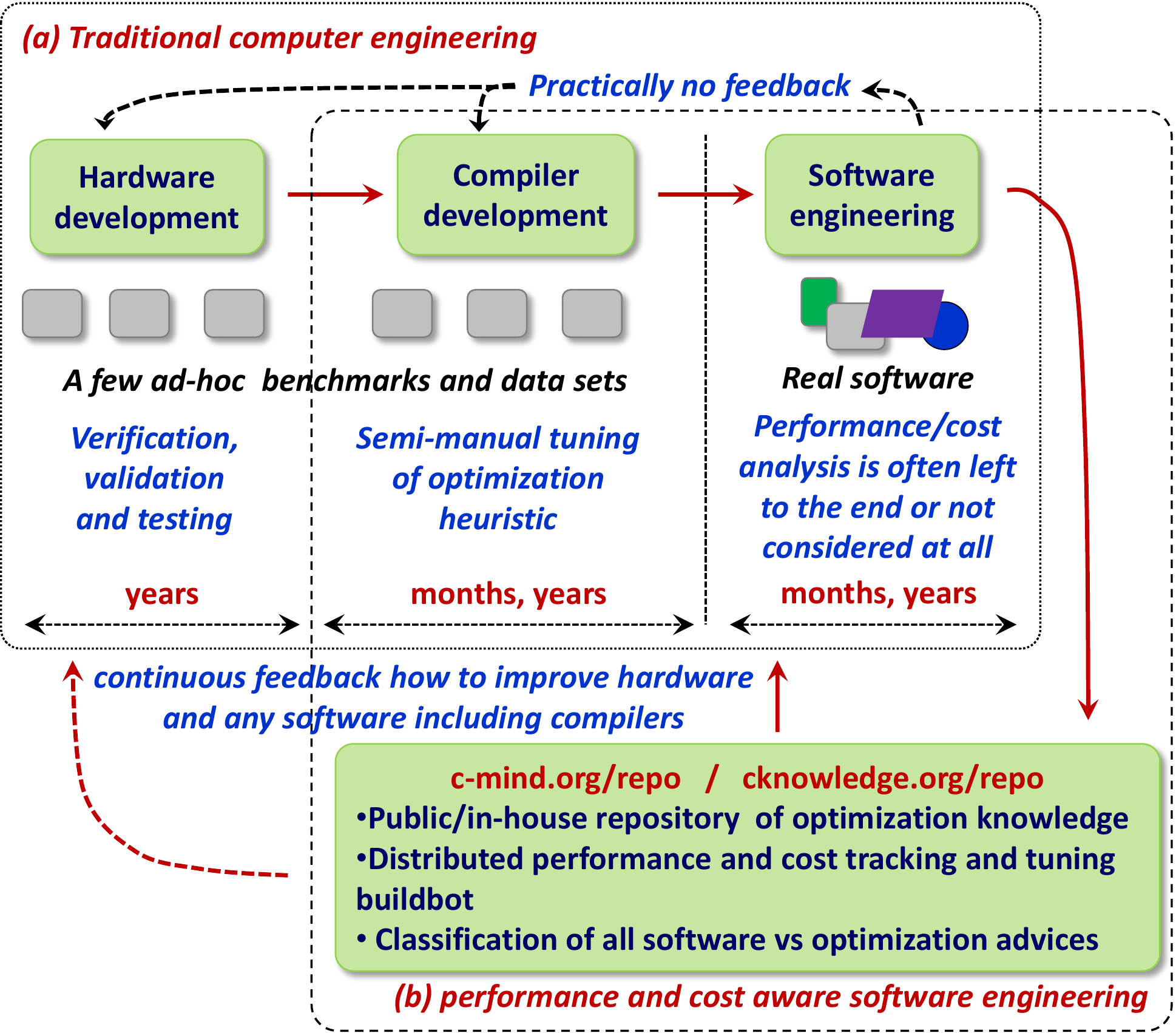}
  \caption{
   (a) traditional computer engineering versus (b) our new
   collaborative performance and cost-aware software/hardware
   co-design as a web service.
  }                                                                                                    
  \label{fig:methodology}
\end{figure}

Indeed, hardware and compiler designers can afford to explore
only a tiny fraction of the whole optimization space together
with a few ad-hoc benchmarks and data sets on a few architectures
when tuning optimization and run-time adaptation
heuristics or training predictive models.
However, this simply can not anymore represent a vast amount
of real-world software and hardware thus explaining why more and
more optimization opportunities are so easily missed.
Furthermore, static compilers often lack data set specific information
and run-time adaptation mechanisms meaning that costly optimization 
process has to be restarted for practically every program input.
Dynamic compilers and run-time autotuning techniques
were not able to solve the problem either.
Such just-in-time frameworks always have a very limited time budget 
to be able to react to varying execution contexts, and hence can not 
explore many aggressive optimizations.
Worse, almost all vast practical optimization knowledge and
experience from the software engineering community during their
own attempts to analyze and tune program behavior is often
wasted instead of helping to improve compilers and run-time
systems as conceptually shown in Figure~\ref{fig:methodology}a.

Rather than waiting until 2020 in a hope that compiler and
hardware designers will somehow find a "holy grail" solution,
we propose an alternative and collaborative solution based on our
interdisciplinary background in physics and AI.
With the help of the software engineering community, we started
gradually identifying real and frequently used software pieces 
ranging from the whole application to jut a few lines of code 
when more than one optimization (or implementation) choice 
is available.
Performance and associated usage costs of such software pieces
usually heavily depend on different optimizations for a given
input (data set), utilized hardware, environment state, and
possible end-user requirements (such as algorithm precision
versus energy usage and performance).

We then propose to use our recent Collective Mind framework and
Hadoop-based repository of knowledge (cM for
short)~\cite{fursin:hal-01054763, software-cm1} to extract and
share such open source software pieces together with various
possible inputs and meta data
at \href{http://c-mind.org/repo}{c-mind.org/repo}.
This meta data is gradually extended by the community via
popular, human readable and easily extensible JSON format~\cite{json-org}
currently describing how to build and run shared pieces together 
with all dependencies on the specific hardware and software 
including compilers, operating systems and run-time libraries.
All these shared software pieces are then continuously and
randomly optimized and executed with different data sets using
distributed cM buildbot for Linux and Windows-based devices~\cite{software-cm1}
or cM node for Android devices~\cite{software-cmn} across shared
computational resources provided by volunteers.
Such resources range from mobile phones, tablets, laptops
and desktops to data centers, supercomputers and cloud services
gradually covering all existing hardware configurations and 
environments.
Furthermore, the community can use light weight cM wrappers
around identified software pieces within a real and possibly
proprietary applications to continuously monitor their behavior
and interactions in the whole software project.
Similar to nature and biological species, such approach treats
all exposed and shared software pieces as \emph{computational
species} while continuously tracking and learning their behavior
versus different optimizations across numerous hardware
configurations, realistic software environments and run-time
conditions.
cM infrastructure then continuously records only the winning
solutions (optimizations for a given data set and hardware) that
minimize all or only monitored costs (execution time, power
consumption, code size, failures, memory and storage footprint,
and optimization time) of a given software piece on a Pareto
frontier~\cite{Kung:1975:FMS:321906.321910} in our public cM repository.

Software engineers can now assemble their projects from the cM
plugins with continuously optimized computational species.
Such software projects can continuously and collaboratively achieve 
better performance while reducing all costs across all hardware 
thus making software engineering performance- and cost-aware.
Furthermore, software developers are now able to practically help
compiler and hardware designers improve their technology
as conceptually shown in Figure~\ref{fig:methodology}b.
Indeed, our approach helps create the first to our knowledge
public, realistic, large, diverse, distributed, evolving and
continuously optimized benchmark with related optimization
knowledge while gradually covering all possible software and
hardware.

At the same time, we can also apply an extensible, top down methodology 
originating from physics when learning behavior of complex systems.
The compiler community first learns and optimizes coarse grain
behavior of large shared software pieces including whole
applications, library functions, kernels and most time consuming
loops versus global compiler optimization flags or other
coarse-grain optimizations.
After enough knowledge is collected, the community can 
gradually move to finer grain levels including just a few source 
lines or binary instructions versus all internal and individual 
compiler optimization decisions via our Interactive
Compilation Interface.
This plugin-based interface is already available in mainline
GCC~\cite{29db2248aba45e59:a31e374796869125}, and we plan to add
it to LLVM in the future~\cite{fursin:hal-01054763}.

More importantly, our approach helps considerably improve 
existing methodology on optimization and run-time adaptation
prediction using machine learning.
Current methodology (used in most of the papers including ours and
referenced at the beginning of this section) usually focuses
on \emph{showing that it is possible to predict} one or several
optimizations to improve execution time, power consumption
or some other characteristics using some off-the-shelf machine
learning techniques such as SVM, (deep) neural networks
or KNN~\cite{citeulike:873540, Hinton06afast, 38115} combined
with a few ad-hoc program or architecture features.
In contrast, our growing, large and diverse benchmark allows 
the community for the first time to apply methodology from sciences 
such as biology, medicine and AI based on big data predictive 
analytics~\cite{DBLP:books/ms/4paradigm09}.
For this purpose, cM infrastructure continuously classifies all
winning species in terms of distinct optimizations and exposes
them to the community in a unified and reproducible way through
the public repository.
This, in turn, allows our colleagues with interdisciplinary
background to help the software engineering community \emph{find best
predictive models} for these optimization classes together 
\emph{with relevant features from software species, hardware, data set and
environment state} either manually or automatically.
Such features (including extraction tool) and predictive models
are continuously added to the species using cM wrappers and their
meta-data thus practically enabling self-tuning software
automatically adaptable to any hardware and environment.

Importantly, cM continues tracking unexpected behavior (abnormal
variation of characteristics of species such as execution time,
or mispredictions from current classification) in a reproducible
way in order to allow the community improve predictive models 
and find missing features that can explain such behavior.
Also, in contrast with using more and more complex and computationally
intensive machine learning techniques to predict optimizations 
such as deep neural networks~\cite{citeulike:873540, Hinton06afast, 38115}, 
we decided to provide a new manual option useful for compiler
and hardware designers.
This option allows the community to combine existing predictive
techniques as a cheap way to quickly analyze large amount
of data, with manually crafted human-readable, simple, compact
and fast rules-based models (decision trees) that can explain and
predict optimizations for a given computational species.
Thus, we are collaboratively building a giant optimization
advice web service that links together all shared software species,
optimizations and hardware configurations while resembling
Wikipedia, IBM Watson advice engine~\cite{FBCFetc10}, Google
knowledge graph~\cite{google-kg} and a brain.

We understand, that the success of our approach will 
depend on the active involvement from the community.
Therefore, we tried to make our approach as simple
and transparent to use as possible.
For example, our light-weight cM version for Android mobile
systems~\cite{software-cmn} is a "one-button approach" allowing
anyone to share their computational resources and tune shared
computational species.
At the same time, extraction of software pieces from large
applications is still semi-manual and may incur some costs.
Therefore we are gradually working on automating this process 
using plugin-based capabilities in GCC and LLVM.
Furthermore, together with participating companies and volunteers,
we already extracted, described and partially~\footnotetext{We can
not share extracted pieces from proprietary software but we still
use them internally for research purposes.} shared 285
computational species together with around 500 input
samples~\footnote{We currently have more than 15000 input samples
collected in our past projects for our shared computational
species~\cite{FCOP2007, midatasets, CHEP2010}. 
However since they require more than
17GB of storage, at the moment we decided to share only
representative ones, i.e. which require distinct compiler
optimization.} from major benchmarks and software projects.
We then validated our approach in STMicroelectronics during 3 months 
to help our colleagues tune their production GCC compiler and 
improve real customer software.
During that time, we continuously optimized execution time, code
size, compilation time and power consumption of all shared computational
species using at least 5000 random combinations of compiler optimization 
flags on spare private cloud servers and mobile phones. 
We also managed to derive 79 distinct optimization optimization
classes covering all shared species (small real applications
or hotspot kernels extracted from large applications with their
run-time data set either manually as we did in~\cite{FCOT2005},
or using Codelet Finder from CAPS Entreprise as we did om the
MILEPOST project~\cite{29db2248aba45e59:a31e374796869125}, or using
semi-manual extraction of OpenCL/CUDA kernels combined with OpenME
plugin interface to extract run-time state~\cite{fursin:hal-01054763}) 
that we correlated with program semantic and dynamic features using
SVM and other predictive analytics techniques.
With the help of domain specialists (compiler engineers), we then
analyzed predictive models for end-user software, found meaningless 
correlations, manually isolated problems~\footnote{In spite
of many papers presents some simple automatic optimization
predictions, our practical and industrial experience with large
data sets shows that it is currently not possible to fully
automate this process. Therefore, manual analysis is still often
required similar to other natural sciences as will be shown later
in this paper.}, prepared and shared counter-example code sample, 
found missing program and input features to fix wrong
classifications, and developed adaptive, self-tuning and statically
compiled code.
Finally, we managed to substitute ad-hoc benchmark used at the
architecture verification department of our industrial partners
with the minimal and realistic one based on derived optimization
classes that helped to dramatically reduce development and
testing time.

These positive outcomes demonstrate how our approach can help
eventually involve the software engineering community into
development and improvement of compilers and hardware.
We also show how continuously growing collective knowledge repository
accessible via unified web service can become an integral part of the 
practical software and hardware co-design of self-tuning computer systems 
while decreasing all development costs and time-to-market for new products.
More importantly, the side effect of our approach to share code and data  
in a reproducible way help support recent international initiatives 
on reproducible research and sustainable software engineering~\cite{cm-reproducibility}.

Our paper is organized as follows. 
This section has introduced the problem, related work and our novel
community-driven approach for performance- and cost-aware
software engineering.
It is followed by Section~\ref{sec:motivation} presenting our
personal and real-life motivating software engineering example
for neural networks with some of the encountered optimization
issues during past 15 years.
Next section \ref{sec:cm} briefly introduces our recent open
source Collective Mind infrastructure and repository to enable
sharing of computational species and collaborative tracking and 
tuning of their performance together with all associated costs
across voluntarily provided computer systems.
Section~\ref{sec:training_set} demonstrates how we
continuously systematize "big performance data" collected by cM,
classify species, and predict optimizations while 
creating a realistic and representative benchmark.
It is followed by Section~\ref{sec:machine_learning} which demonstrates
how to understand machine learning and improve optimization predictions.
Section~\ref{sec:features} demonstrates how to find missing features 
to explain unexpected behavior of computational species and improve 
optimization predictions.
It also shows how to build adaptive and self-tuning applications 
assembled from available computational species as plugins.
Finally, we conclude paper and describe future research
and development directions in Section~\ref{sec:conclusion}.

\section{Real-life motivating example}
\label{sec:motivation}

\begin{figure}[htb]
  \centering
   \includegraphics[width=3.8in]{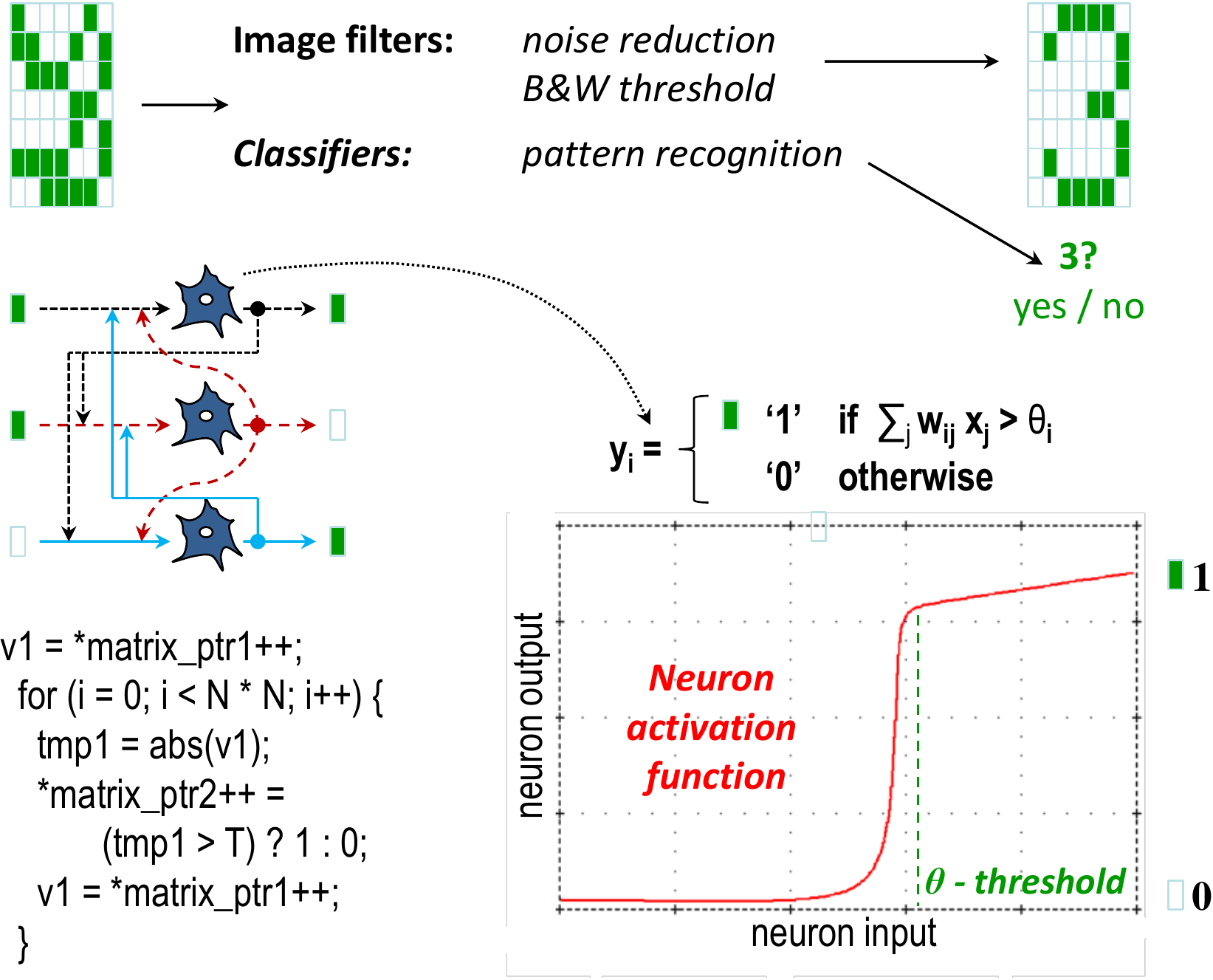}
  \caption{
  Conceptual example of pattern recognition, image filtering
  and character noise reduction using Hopfield fully
  interconnected and recurrent neural network. Simple C kernel
  is a part of a neuron activation function processing thresholds
  for all neurons.
  }
  \label{fig:neural-network}
\end{figure}

One of the authors original research started more than twenty years ago
was to develop and analyze various artificial neural networks as a part
of a possible non-traditional and brain-inspired computer.
Such networks can mimic brain functions and are often used for  
machine learning and data mining~\cite{citeulike:873540}. 
For example, Figure~\ref{fig:neural-network} shows one of the oldest
and well-known one-layer, fully interconnected, recurrent 
(with feedback connections) Hopfield neural network~\cite{Hopfield01041982}.
It is a popular choice for function modeling, pattern recognition
and image filtering tasks including noise reduction.
Implemented as a software, this neural network has a fairly simple
and regular code where each neuron receives a weighted sum of all inputs
of an image as well as outputs of all other neurons. 
This sum is then processed using some neuron activation 
function including sigmoid or linear ones to calculate the output value.
The small and simple C kernel presented in above
Figure~\ref{fig:neural-network} is one of many possible
implementations of a threshold filter we used as a part of 
a linear activation function, i.e. switching neuron output 
from 0 to 1 when its input meets a given threshold.

Very simplistically, the quality of a neural network is usually
determined by its processing speed as well as capacity (maximum
amount of patterns or information that can be stored in such
networks) and recognition accuracy (correct predictions versus
failures).
It heavily depends on the total number of neurons, connections
and layers~\cite{citeulike:3735200}, and is primarily limited by 
the speed and resources of the available hardware including
specialized accelerators.
Hence, neural network software/hardware co-design process 
always involves careful balancing of performance versus
all associated costs including storage size, memory footprint, 
energy consumption, development time and hardware price 
depending on usage scenarios and required time to market.
Indeed, our research on improving neural networks requires 
many iterative runs of a slightly evolving modeling software 
with varying parameters to maximize prediction accuracy. 
In this case, our main concern is about minimizing compilation
and execution time of each execution across available hardware.
However, when the best found network is found and deployed in
a large data center or cloud service (for example, for big data
analysis), end users would like to minimize all
additional costs including energy and storage consumption across
all provided computer systems.
Finally, when deploying neural networks in small, autonomic and
possibly mass-produced devices such as surveillance cameras and
mobile phones or future robots and Internet of Things objects,
more strict requirements are placed on software and hardware
size, memory footprint, real time processing, and the cost of the
whole system.

Twenty years ago, our software engineering of neural networks
was relatively straightforward.
We did not have a choice but to simply select the latest hardware
with the accompanying and highly tuned compiler to achieve nearly
peak performance for our software
including for the code shown in Figure~\ref{fig:neural-network}.
Therefore, in order to innovate and process more neurons
and their configurations, we usually had to wait for more
than a year until arrival of a new hardware.
This hardware would likely double performance of our software and
provide more memory and permanent storage but often at a cost
of higher power consumption and thus dramatically rising
electricity bill.

\begin{figure}[htb]
  \centering
   \includegraphics[width=4.5in]{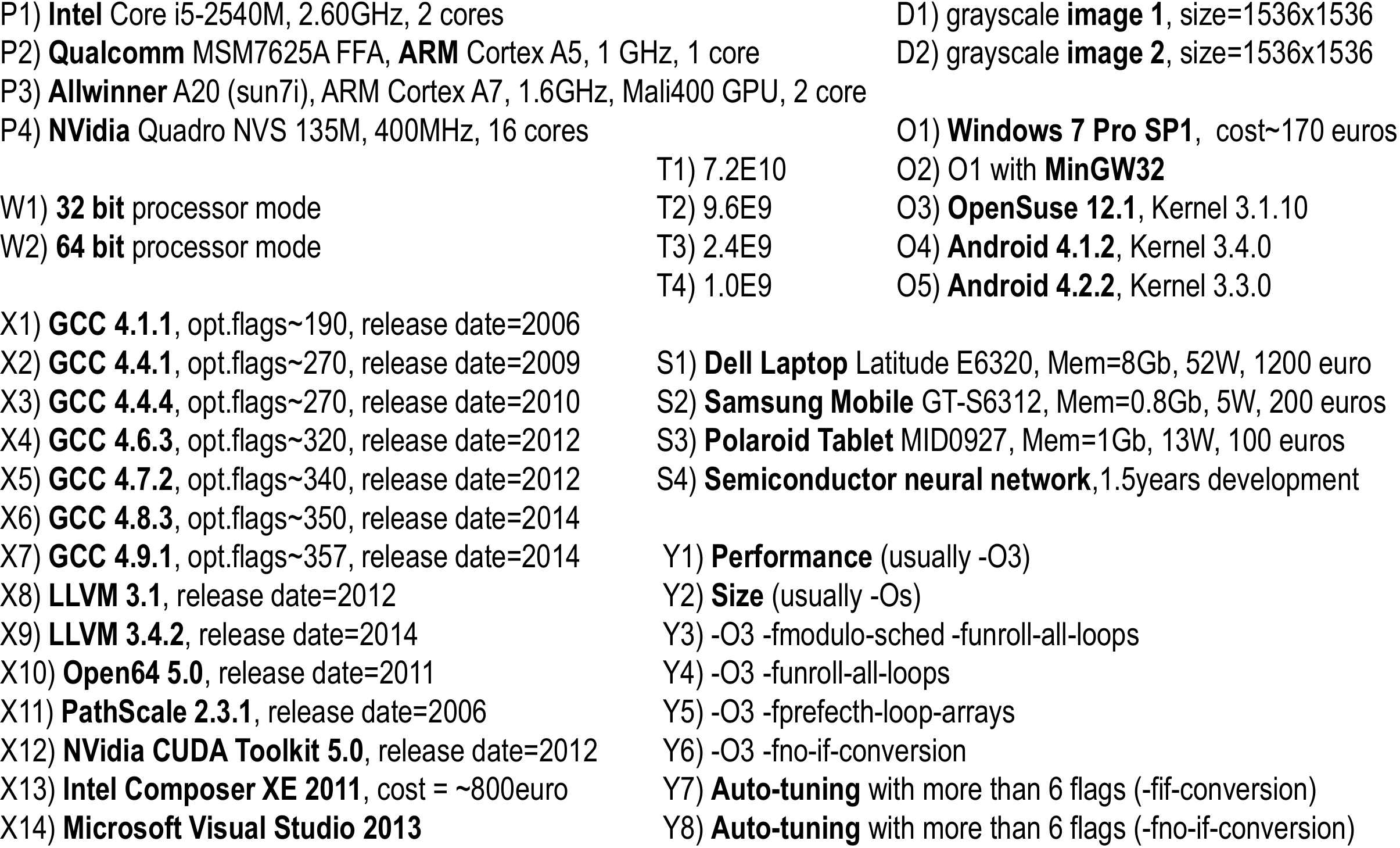}
(a) \\

   \includegraphics[width=4.5in]{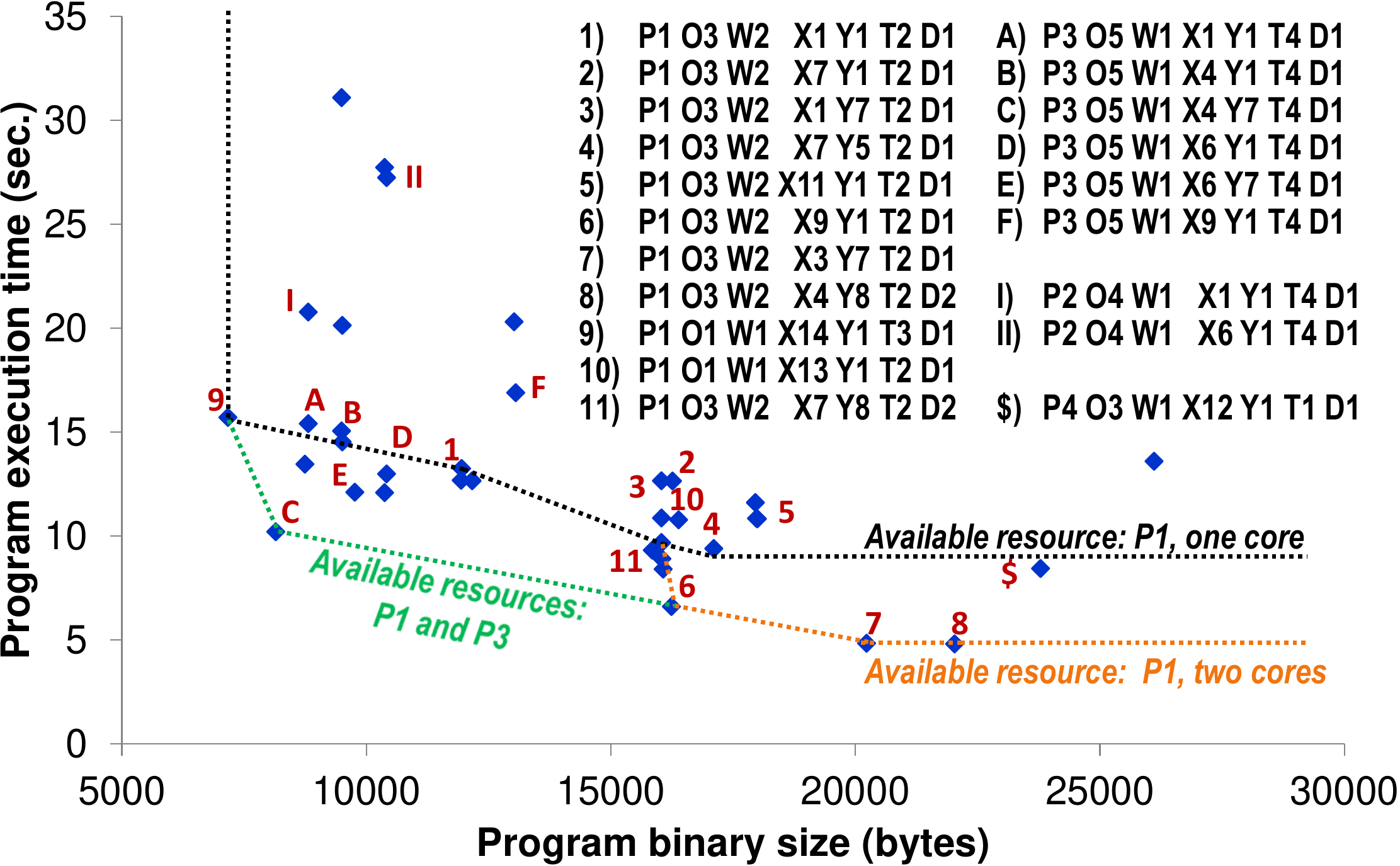}
(b)
   \caption{ 
   (a) A small subset of various hardware, software, development
   tools and optimizations used in our research on neural
   networks in the past 20 years (P - processors, W - processor
   mode, X - compiler, O - operating system, S - system, T - total
   number of processed pixels or neurons, D - software data set,
   Y - compiler optimization used)
   (b) 2D projection of the multidimensional space of characteristics
   together with winning solutions on the Pareto frontier (all
   data and interactive graphs are available at 
   \href{http://c-mind.org/nnet-tuning-motivation}{c-mind.org/nnet-tuning-motivation}).
   }
  \label{fig:motivation}
\end{figure}

In contrast, we now have an impressive choice of hardware 
of all flavors which our software can be executed on.
Each year, there are numerous variations of processors appearing
on the market with different features (properties) including
frequency, number of cores, cache size, ISA extensions,
specialized hardware accelerators (such as GPU and even revived
semiconductor neural networks), power consumption and price.
Furthermore, we can now have an easy access to large-scale parallel 
resources from home via gaining popularity virtualized cloud services
from Amazon, Google, Microsoft and others.
Therefore, the number of experiments we can now run is mainly
limited by the price we can afford to pay for computing 
services.
At the same time, we also enjoy continuous community-driven
improvements of operating systems together with numerous free
or proprietary libraries and software development tools including
popular optimizing compilers such as GCC and LLVM.
One may expect that with so many advances in the computer
technology, practically any recent compiler would generate the
fastest and most energy efficient code for
such an old, simple, small and frequently used software piece
shown in Figure~\ref{fig:neural-network} across existing hardware.
Nevertheless, since we pay for our experiments, we eventually 
decided to validate their performance/cost efficiency.

For the sake of accountability and reproducibility, we started
gradually collecting at
\href{http://c-mind.org/nnet-tuning-motivation}{c-mind.org/nnet-tuning-motivation}
various information about several computer systems we used including
their price, cost, available operating systems, compilers and
optimizations.
Figure~\ref{fig:motivation}a shows a tiny subset of this
multidimensional space of design and optimization choices.
At the same time, whenever running real experiments, we also
started recording their execution time~\footnote{In this paper,
for the sake of simplicity and without loosing generality, when
speaking about performance tuning, we mean reducing execution
time. However, on modern out-of-order processors with complex
memory hierarchy, the dependency between performance (speed
of execution) and total execution time may be non-linear. Thus,
depending on user requirements, these characteristics have to be
tuned separately.} and all associated costs including compilation
time, code size, energy usage, software/hardware price and
utility bill.
Furthermore, we decided to perform a simple and well-known
optimization compiler flag autotuning~\cite{acovea,
29db2248aba45e59:a31e374796869125} with at least 100 iterations
to see whether there is still room for improvement over 
the fastest default compiler optimization level (-O3).
Figure~\ref{fig:motivation}b shows one of many possible
2D projections of the multidimensional space of characteristics
(which we consider as costs of running our experiments or tasks).
We then gradually track the winning solutions that maximize
performance and at the same time minimize all costs using our
experience in physics and electronics, namely by applying Pareto
frontier filter~\cite{Kung:1975:FMS:321906.321910}.

We quickly realized that in contrast to the traditional
wisdom, the latest technology is not necessarily the fastest
or most energy efficient and further optimization is always
required.
For example, when moving from GCC 4.1.1 (released in 2006) to GCC 4.9.1 (released
in 2014), we observed
a modest 4\% improvement~\footnote{Similar to physics, we execute optimized
code many times, check distribution of characteristics for 
normality~\cite{EPL1983}, and report expected value if variation
is less than 3\%} in single core execution time of our neural 
network and 2\% degradation in a code size on Intel E6320 based system (released in 2008). 
However, 8 years old GCC 4.1.1 
can achieve 27\% improvement in execution time after autotuning
(which comes at cost of 100 recompilations and executions as well as
increasing binary size by 34\%)! 
Interestingly, 8 years old PathScale 2.3.1 produces faster code
than the latest version of GCC 4.9.1 and LLVM 3.4.2!
Furthermore, when using internal parallelization, LLVM 3.4.2 beats GCC 4.9.1
by about 23\% but has a sub-linear scaling versus number of threads.
In contrast, 2 years old GCC 4.6.3 achieves the best result and linear scaling
versus number of threads when using both parallelization and autotuning!

\begin{figure}[htb]
  \centering
   \includegraphics[width=5.5in]{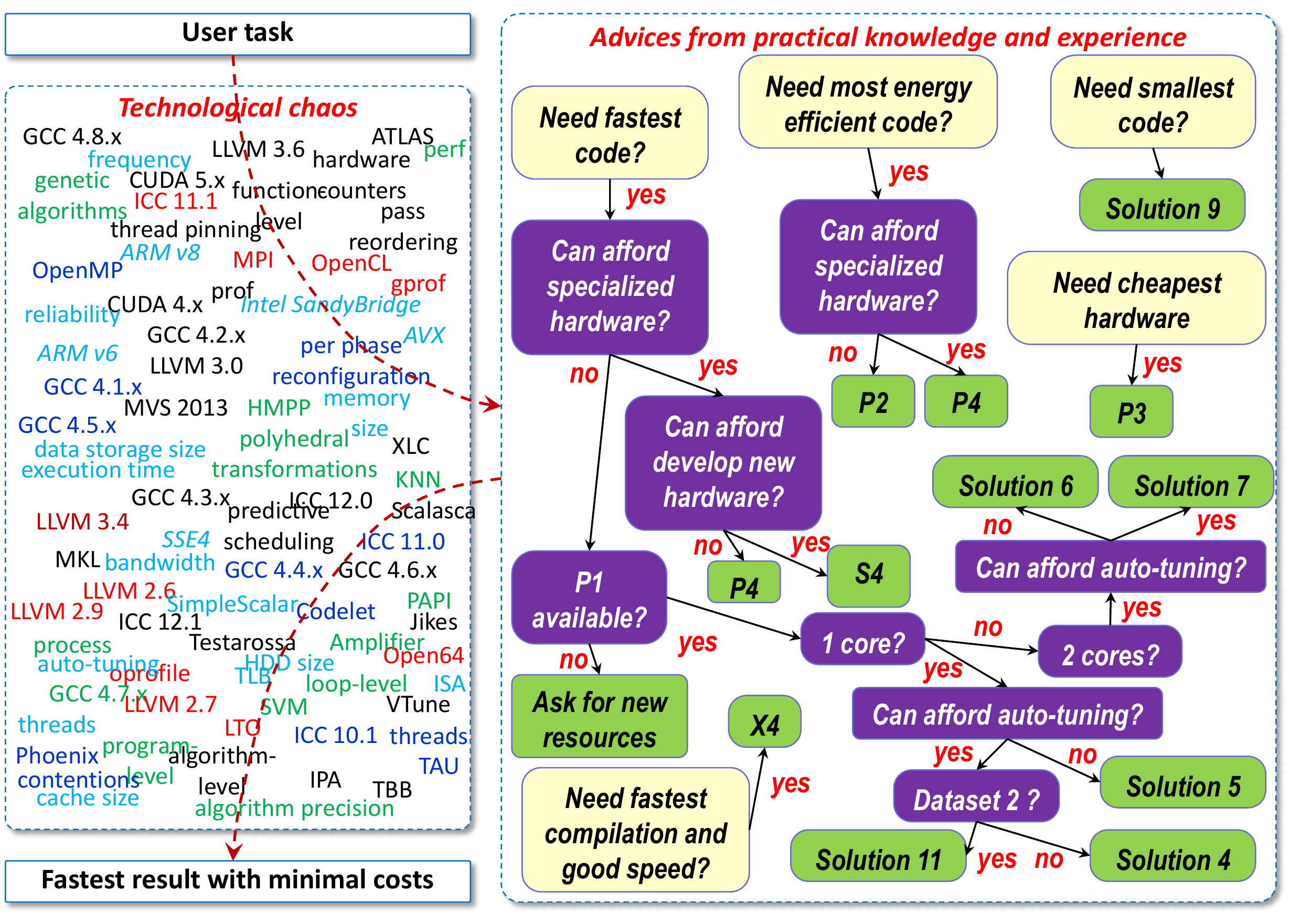}
  \caption{ 
  Example of gradually and manually crafted advices as decision trees
  to deliver best performance and cost for our neural network
  depending on available resources, usage scenarios (requirements) 
  and data sets.
  }
  \label{fig:my-decision-tree}
\end{figure}

When running the same code on cheap, commodity mobile phones with ARM
architecture, the execution time increased dramatically by around 5 times!
However, the power consumption dropped by about 10 times!
When trying to use specialized hardware (GPUs or our semiconductor neural
networks), we could increase execution time by about tens to hundreds of times, 
but at a considerable development cost and time to market.
Furthermore, with time, we discovered that the same best found optimization
for one class of images can considerably degrade performance on another
class of images (as we explain in Section~\ref{sec:features}).
We also encountered problems with cache contentions on multi-core systems,
sub-linear scaling on many core systems, unexpected frequency scaling,
nondeterministic IO for large images, and many other problems that
had to be addressed by new optimizations.
These issues can not be easily solved by static compilers
due to a fundamental problem of a lack of run-time information
at compile time.
Therefore, we even tried to move to dynamic and possibly adaptive
languages including Java and Python but were not yet able
to achieve similar performance while spending even more energy
and storage during just-in-time compilation.

Sadly and similar to many other scientists and software engineers,
we now have to waste considerable amount
of our time on a tedious and ad-hoc navigation through the
current technological chaos to find some good
hardware, software and optimization solutions that can speed up our
programs and reduce costs instead of innovating as conceptually summarized in Figure~\ref{fig:my-decision-tree}.
Worse, software engineers are often not even aware of all
available design and optimization choices they have to improve
performance of their software and reduce development and usage
costs.
Furthermore, costs that has to be minimized depend on usage
scenarios: in mobile systems running out of battery, one may want to fix 
a power budget and then balance execution time and algorithm accuracy; 
in embedded devices, code size and consumed energy
may be more important that execution time; JIT may require careful
balancing of compilation and optimization times versus potential 
performance gains, while users of data centers and supercomputers 
may care primarily about both execution time and the price of computation.
Therefore, we strongly believe that current performance- and
cost-blind software engineering has be changed to improve
productivity and boost innovation in science and technology.

\section{Public and Open Source Collective Mind Infrastructure and Repository}
\label{sec:cm}

\begin{figure}[htb]
  \centering
   \includegraphics[width=5.5in]{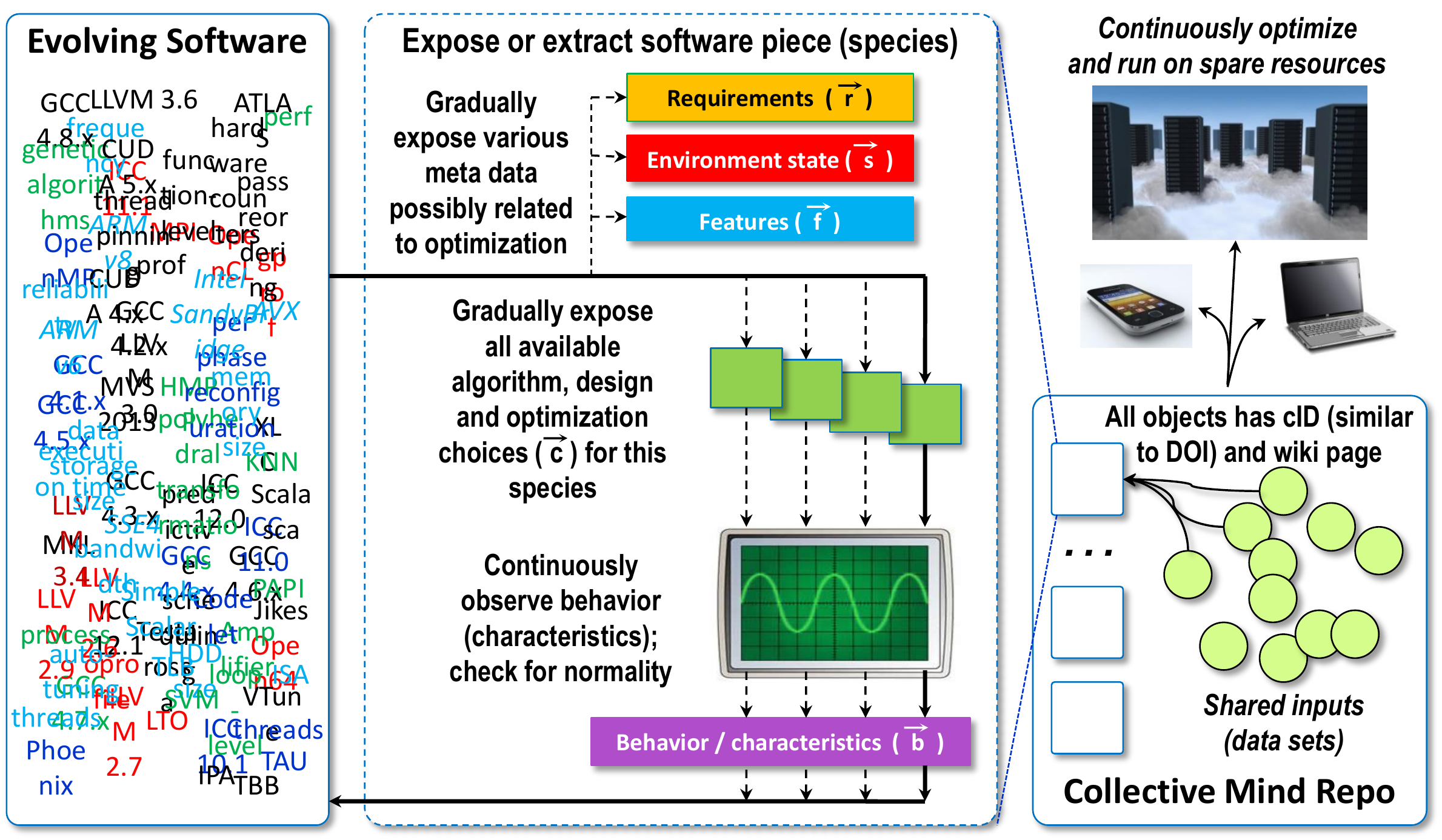}
   \caption{ 
    Collective Mind Framework and Repository (cM) help
    to decompose any complex software into pieces with
    light-weight wrappers that expose design and optimization
    choices, measured characteristics, features and environment
    state in a unified and mathematical way. It was developed
    to unify and systematize software autotuning, make
    it practical and reproducible, and distribute it among
    numerous computing resources such as mobile phones and data
    centers shared by volunteers~\cite{fursin:hal-01054763,
    software-cm1}.
   }
  \label{fig:cm}
\end{figure}

Eventually, we started searching for a possible solution that
could liberate software developers from the tedious and not
necessarily relevant job of continuous optimization and
accounting while gradually making existing software 
performance- and cost-aware.
At first, we tried to create a simple database of optimizations 
and connect it to some existing benchmarking and autotuning tools 
to keep track of all optimizations~\cite{Fur2009,29db2248aba45e59:a31e374796869125}.
However, when trying to implement it within production environments
of our industrial partners, we faced several severe problems including 
difficulty to expose all design and optimization choices from continuously
evolving software, and difficulty to reproduce performance numbers
collected from different machines.
This eventually pushed us to develop a full-fledged repository
of knowledge with unified web services (Collective Mind or cM for short) similar to ones that helped successfully 
systematize research and experimentation in biology, genomics and other 
natural sciences.
Such repository should be able to keep the whole autotuning
setups with all dependencies including optimized software, data
sets and autotuning tools.
This, in turn, should allow us to distribute the whole
autotuning setups among many users to crowdsource software
optimization (or any other experimentation) in a reproducible way while considerably reducing
usage costs.

Briefly~\footnote{Though we provide minimal information 
about Collective Mind framework in this paper, it should be enough 
to understand proposed concepts. However, in case of further interest,
more details can be found in~\cite{fursin:hal-01054763,software-cm1}},
cM helps decompose software into standalone pieces interconnected 
through cM wrappers.
Such light-weight wrappers currently support major languages
including C, C++, Fortran, Python, PHP and Java, and allow the
community to gradually expose various design and optimization
choices~\textbf{c}, features~\textbf{f}, dependencies on other
software and hardware, monitored characteristics
(costs)~\textbf{b} and environment state~\textbf{s} in a unified
way through extensible JSON format~\cite{json-org}.
This allowed us to formalize almost all existing autotuning
techniques as finding a function of a behavior of a given software
piece~\emph{B} running on a given computer system with a given
data set, selected hardware design and software optimization choices~\textbf{c},
and a system state~\textbf{s} (\cite{fursin:hal-01054763}):
\begin{displaymath} 
\bf{b} = \emph{B}(\bf{c},\bf{s})
\end{displaymath}
Furthermore, software pieces can be extracted and then shared together with their 
wrappers and data sets samples in the Hadoop-enabled~\cite{software-elastic-search} cM repository.
For example, with the help of our colleagues and supporters,
we already gradually and semi-automatically extracted and shared 285
software pieces together with several thousand data set pairs
from several real software projects as well as 8 popular benchmark suits including 
NAS, MiBench, SPEC2000, SPEC2006, Powerstone, UTDSP and SNU-RT.
This can liberate software engineers from developing their own ad-hoc and 
complex tuning setups in favor of implementing common autotuning pipelines
consisting of shared software pieces, data sets, tools and optimization
space exploration modules.
Such pipelines can then be easily shared and distributed across a 
large number of diverse computer systems either using open source
\emph{cM buildbot} or a small \emph{cM node} that can deploy
experiments on Android-based devices~\cite{software-cmn}.
cM will then continuously "crawl" for better optimizations for
all shared software pieces, data sets and compilers, while
recording experiments in a reproducible way in the public
cM repository \href{http://c-mind.org/repo}{c-mind.org/repo}.

As requested by our industrial partners, we first used cM to
implement compiler optimization flag autotuning pipeline.
Though this pipeline continues growing and now supports 
other optimizations including polyhedral source-to-source transformations, 
MPI/OpenMP tuning and fine-grain compiler transformations 
as can be seen at \href{http://c-mind.org/ctuning-pipeline}{c-mind.org/ctuning-pipeline},
compiler flag optimization problem remains critical for new hardware
and unsolved for several decades due to growing optimization spaces.
Indeed, latest GCC, LLVM and Intel compilers feature hundreds of 
optimizations which have to be carefully orchestrated for all 
existing software and hardware.
At the same time, it should be simple enough to demonstrate the
concepts of our approach which follows top-down optimization methodology.
Therefore, we start from coarse-grain optimizations and gradually move to 
finer-grain levels motivated by our experience in physics when gradually
moving from three Newton laws to quantum mechanics.

{\center
\begin{table}
{\center
\begin{tabular}{|p{6.6in}|}
 \hline
 -O3 -fif-conversion \textit{\textbf{-fno-ALL}} \\
 -O3 --param max-inline-insns-auto=88 -finline-functions \textit{\textbf{-fno-ALL}} \\
 -O3 -fregmove -ftree-vrp \textit{\textbf{-fno-ALL}} \\
 -O3 -fomit-frame-pointer -fpeel-loops -ftree-fre \textit{\textbf{-fno-ALL}} \\
 -O3 -falign-functions -fomit-frame-pointer -ftree-ch \textit{\textbf{-fno-ALL}} \\
 -O3 -ftree-dominator-opts -ftree-loop-optimize -funswitch-loops \textit{\textbf{-fno-ALL}} \\
 -O3 -ftree-ccp -ftree-forwprop -ftree-fre -ftree-loop-optimize \textit{\textbf{-fno-ALL}} \\
 -O3 -finline-functions -fivopts -fprefetch-loop-arrays -ftree-loop-optimize -ftree-vrp \textit{\textbf{-fno-ALL}} \\
 -O3 -fdce -fgcse -fomit-frame-pointer -freorder-blocks-and-partition -ftree-reassoc -funroll-all-loops \textit{\textbf{-fno-ALL}} \\
 -O3 -fivopts -fprefetch-loop-arrays -fsched-last-insn-heuristic -fschedule-insns2 -ftree-loop-optimize -ftree-reassoc -ftree-ter \textit{\textbf{-fno-ALL}} \\
 -O3 -fforward-propagate -fguess-branch-probability -fivopts -fmove-loop-invariants -freorder-blocks -ftree-ccp -ftree-ch -ftree-dominator-opts -ftree-loop-optimize -ftree-reassoc -ftree-ter -ftree-vrp -funroll-all-loops -funswitch-loops -fweb \textit{\textbf{-fno-ALL}} \\
 \hline
\end{tabular}
}
 \caption{Example of a few top ranked optimization classes for GCC 4.6.3
 across Intel E5520 based computer systems from the French
 GRID5000 public data center~\cite{GRID5000} using cM buildbot
 with all shared software pieces and data sets.}
\label{tab:pruned}
\end{table}
}

Our autotuning pipeline randomly selects a software piece from
all shared ones, builds it with randomly generated combinations
of compiler optimization flags of format \emph{-O3
-f(no-)optimization\_flag --parameter param
= random\_number\_from\_range}, and runs it on a randomly selected
spare computer system with a randomly selected shared data set.
Measured characteristics (costs) are then processed using Pareto
frontier filter to record or update the winning solution for
a given software piece (optimization versus hardware, compiler 
and data set) in the cM repository.
For example, table~\ref{tab:pruned} presents some of the winning 
combinations of flags for GCC 4.6.3 used in the production 
environment by one of our industrial partners. 
Note that meta flag~\textit{\textbf{-fno-ALL}} means that
all other optimization flags have been gradually switched off to 
leave only the influential ones and thus help compiler designers
understand missed optimization opportunities.
This allows developers to continuously monitor and reuse
best solutions for shared software pieces through unified cM web
services depending on their further intended usage.
For example, the fastest and energy efficient solution is often used 
for HPC systems, the smallest - for computer
systems with very limited resources such as credit card chips
or the future "Internet of Things" devices, or balanced for both
speed and size when used in phones, tablets and other mobile devices.

\section{Classifying computational species}
\label{sec:training_set}

Though cM helped simplify, unify, automate, speed
up and preserve a tedious process of hardware benchmarking and 
software autotuning, it also encountered a new serious problem.
Even with a modest number of 400 volunteers participated in our
project since the beginning of 2014 to crowdsource compiler
autotuning using our Android-based cM node~\cite{software-cmn},
we immediately faced a big data problem - too much collected 
statistics that our repository and infrastructure can process.
However, we have already faced a similar problem in natural sciences and AI for
many years and managed to effectively tackle it using predictive
analytics (statistical analysis, data mining and machine
learning)~\cite{citeulike:873540, Hinton06afast, 38115,
DBLP:books/ms/4paradigm09}.
For example, we used to classify numerous physical objects
or biological species in terms of behavior and features while
leaving only representative classes and thus considerably reducing
analysis time and required storage.

We propose a similar approach for software engineering.
Collective Mind Framework, Repository and mathematical
formalization allows us to treat all shared software pieces
as computational species running across numerous hardware 
configurations.
Naturally, these species will behave differently depending
on their data sets, hardware used and environment state (which
includes the state of the species, Operating System and hardware
as well as interaction with other species).
Since our main goal is to help software engineers find better
optimizations that improve overall behavior of their software 
pieces under all usage conditions (reduce execution time and all other
costs), we implemented continuous clustering of all available
species that share the same top performing optimizations
such as ones shown in Table~\ref{tab:pruned}.
Our expectation is that software pieces belonging to the same
cluster, i.e. share similar optimizations, will also share some 
features describing their semantics, data set and run-time behavior.
This, in turn, would allow us to automatically classify
previously unseen computational species from the community and relate them 
to existing optimization cluster based on their features
thus effecitvely predicting optimizations and dramatically reducing tuning 
time and storage size.

However, the major challenge with such approaches is to assemble a large 
and diverse enough training set, and to find meaningful features that can
correctly separate optimization clusters.
Collective Mind Repository helps solve the first problem
by collecting a large number of real and diverse software pieces
from the community.
At the same time, from our practical experience in using machine
learning for program optimization and hardware
designs~\cite{29db2248aba45e59:a31e374796869125, DJBP2009},
we believe that it is not currently possible to fully automate
this process due to a practically infinite feature space.
Instead, we propose to use a collaborative approach similar
to ones currently used in many natural sciences.
We expose optimization clusters
together with software species and data sets through our public cM repository 
in a reproducible way to let the interdisciplinary community find 
and share meaningful features and correlations similar to ones 
shown in Figure~\ref{fig:my-decision-tree}.
This approach contrasts with some existing works on machine
learning for compilation and architecture.
They were referenced in the related work section and mainly focus on
showing that machine learning can be used to predict optimizations while using just a few
benchmarks, data sets and ad-hoc features.

\begin{figure}[htb]
  \centering
  \includegraphics[width=6.5in]{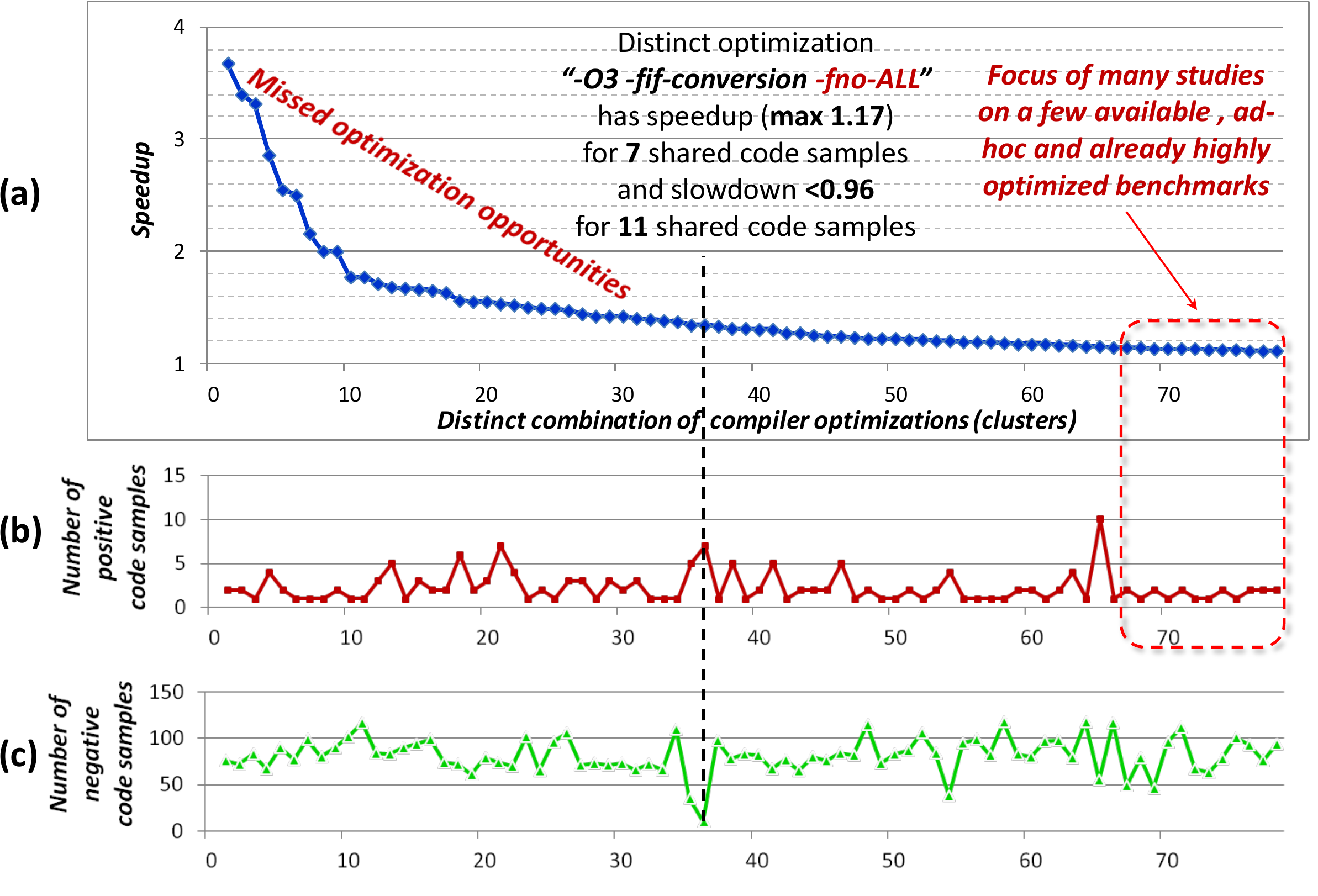}
\caption{(a) 79 distinct combinations of optimizations
(optimization clusters) covering all 285 shared software species and data set
samples on Intel E5520, GCC 4.6.3 and at least 5000 random
combinations of optimization flags together with maximum speedup achieved
within each optimization cluster; (b) number of benchmarks with
speedup at least more than 1.1 for a given cluster; (c) number
of benchmarks with speedup less than 0.96 (slowdown) for a given
cluster.}
  \label{fig:iterative-speedups}
\end{figure}

Our approach helped our industrial partners tune and cluster all
shared software species across several production compilers and
architectures.
Next, we present a small subset of our experimental results 
for GCC 4.6.3 and Intel E5520. 
All other results are continuously updated at our public
cM repository at \href{http://c-mind.org/repo}{c-mind.org/repo}.
By now, cM has found a pool of 79 distinct combinations of optimization 
flags (clusters) for this compiler covering all shared software pieces 
and data set samples. 
Figure~\ref{fig:iterative-speedups} shows maximum speedups
achieved for each top performing and representative optimization
across all benchmarks together with the number of benchmarks
which achieve highest speedup using this optimization (or at
least more than 1.1 ) and the number of benchmarks with speedups
less than 0.96 (slowdown) for the same optimization. 
For example, distinct combination of optimizations \emph{-O3 -fif-conversion
-fno-ALL} achieved maximum speedup on 7 benchmarks (including
1.17 speedup on at least one of these benchmarks) and slowdowns
for 11 benchmarks. 
In case some optimization cluster does not have slowdowns across
all shared software pieces, it can help automatically substitute 
a best and often manually crafted default compiler optimization 
level such as -O3, thus practically helping to improve existing compilers.

\section{Understanding and improving machine learning to predict optimizations}
\label{sec:machine_learning}

\begin{table}
\center
\begin{tabular}{|p{1.5in}|p{1.5in}|}
 \hline
 \textbf{Number of species} & \textbf{Prediction accuracy} \\
 \hline
 12 from prior work~\cite{29db2248aba45e59:a31e374796869125} & 87\% \\
 \hline
 285 from current work & 56\% \\
 \hline
\end{tabular}
\caption{Prediction accuracy when using optimized SVM with full
cross-validation for 12 and 285 software species together with all available 
semantic and dynamic features from MILEPOST GCC and hardware counters respectively.}
\label{tab:predictions}
\end{table}
As the first approximation, we decided to reuse existing machine
learning techniques and build a predictive model (classifier)
that can associate any previously unseen species with a unique
optimization cluster based on some software features.
We also decided to reuse and validate already
available semantic and dynamic features from our previous work 
on machine learning based compiler~\cite{29db2248aba45e59:a31e374796869125}.
For this purpose, we generated and shared in the cM repository
a feature vector \textbf{f} for each software species using
56 semantic program features from the MILEPOST
GCC~\cite{29db2248aba45e59:a31e374796869125} (extracted during
compilation at \emph{-O1} optimization level after \emph{pre}
pass)). In addition, we collected 29 following dynamic features (hardware counters collected
by default using standard performance monitoring tool \emph{perf}
when running unoptimized software piece on a shared computing
resource with Linux-based OS):
 {\small \textit{"cycles",
"instructions",
"cache-references",
"cache-misses",
"L1-dcache-loads",
"L1-dcache-load-misses",
"L1-dcache-prefetches",
"L1-dcache-prefetch-misses",
"LLC-prefetches",
"LLC-prefetch-misses",
"dTLB-stores",
"dTLB-store-misses",
"branches",
"branch-misses",
"bus-cycles",
"L1-dcache-stores",
"L1-dcache-store-misses",
"L1-icache-loads",
"L1-icache-load-misses",
"LLC-loads",
"LLC-load-misses",
"LLC-stores",
"LLC-store-misses",
"dTLB-loads",
"dTLB-load-misses",
"iTLB-loads",
"iTLB-load-misses",
"branch-loads",
"branch-load-misses"
}}.

We then passed 79 optimization clusters, 86 features and either all 285 shared
species or a small subset of 12 ones used in some of our past
works through a standard SVM classifier from R package~\cite{citeulike:873540, r-project} 
with full cross-validation as described in~\cite{29db2248aba45e59:a31e374796869125, CFAP2007}.

\begin{figure}[htb]
  \centering
  \includegraphics[width=5.6in]{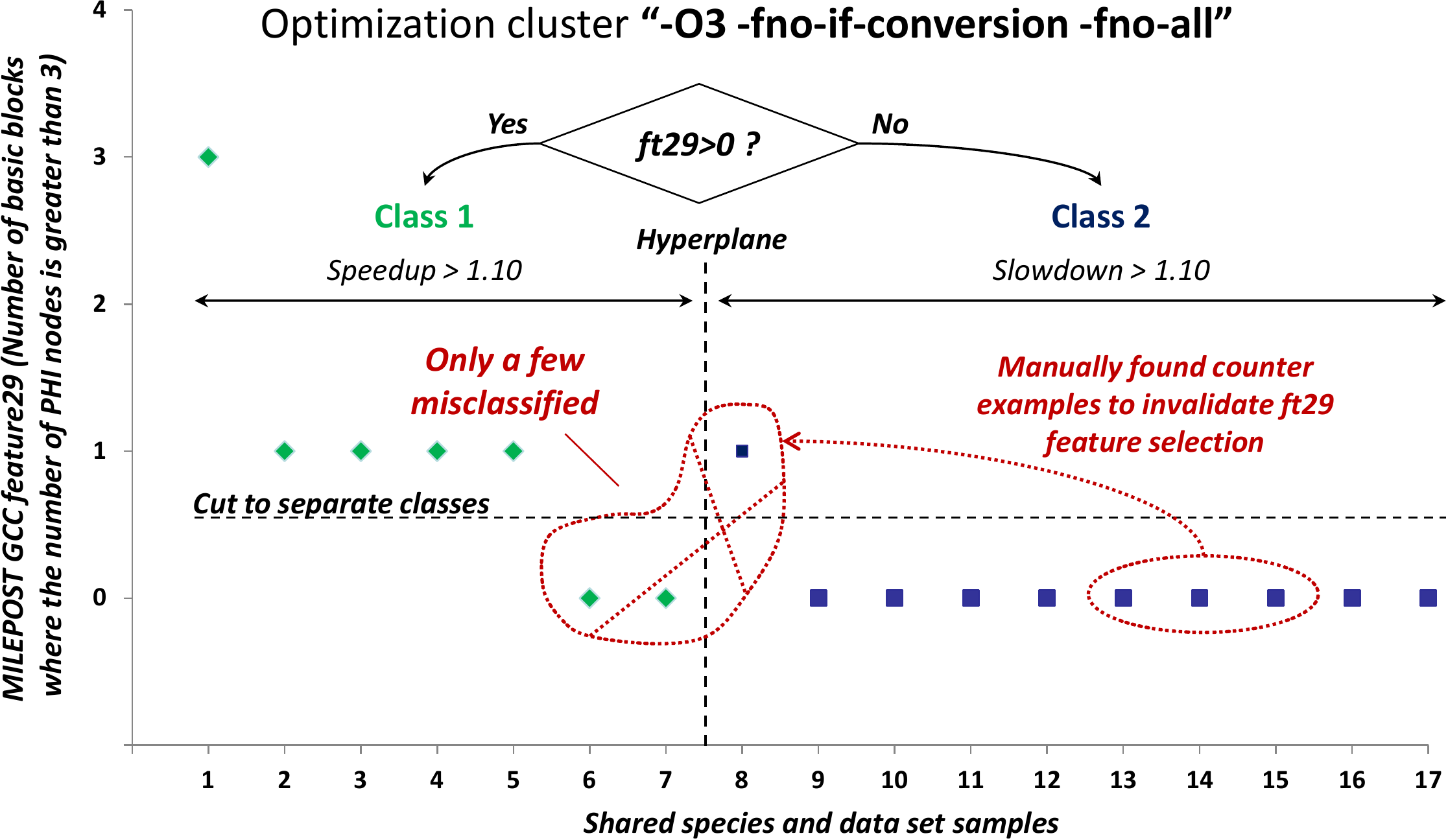}
\caption{Visualization of the predictive model built by SVM to predict optimization 
cluster "-O3 -fno-if-conversion -fno-ALL" using MILEPOST GCC features. We provided 
several counter examples to show that automatically found correlation can be totally misleading.}
  \label{fig:feature-space}
\end{figure}
Results shown in Table~\ref{tab:predictions} demonstrate that
we can obtain a relatively high prediction accuracy when using
just a few software species thus supporting findings from similar
works on machine learning for autotuning, i.e. that it is possible
to predict optimizations using machine learning.
However, with a new opportunity provided by cM to use a much
larger training set of shared software species from the community, prediction
accuracy dramatically dropped exhibiting close to random behavior
(50\%), i.e. with no meaningful predictions.
At the same time, Collective Mind approach can now help understand 
such a variation in prediction accuracy by exposing
all optimization and modeling results in a reproducible way
to experts. 
Our colleagues and compiler specialists from STMicroelectronics 
analyzed one of the simplest optimization clusters \emph{-O3 -fif-conversion -fno-ALL} 
which has a relatively high prediction rate when using a few or all shared programs, 
i.e. only 3 mispredictions out of a pool of 17 optimization clusters (7 positive speedups 
and 10 negative ones) as shown in Figure~\ref{fig:iterative-speedups}. 
We then incrementally removed all unrelated features that did not
influence predictions leaving only one semantic feature from
MILEPOST GCC (ft29) that counts the number of basic blocks where the
number of phi-nodes is greater than 3.
Visualization of the predictive model at Figure~\ref{fig:feature-space}
shows that SVM derived a decision ft29 $\rangle$ 0 to effectively
separate two classes with only 3 mispredictions out of 17.

At this stage, many existing academic works will conclude that relevant
feature is found and it is possible to use machine learning
to predict optimization.
However, compiler or hardware designers also need to understand whether this feature
makes sense in order to improve their technology.
Since our colleagues did not manage to explain this feature,
they decided to try to find a counter example to invalidate this correlation.
Therefore, they selected a simple \emph{blocksort function} from the shared \emph{bzip2} species 
that has 0 phi-nodes and tried to manually add phi-nodes by converting source code 
as following (added lines are highlighted):

{
\noindent
  \hphantom{40}  \textbf{volatile int sum, value = 3;} \\
  \hphantom{40}  \textbf{int sumA = 0; int sumB = 0; int sumC = 0;} \\
  \hphantom{40}  for (j = ftab[ss $\langle$$\langle$ 8] \& (~((1 $\langle$$\langle$ 21))) ; j $\langle$ copyStart[ss] ; j++) \{\\
 \hphantom{140}      k = ptr[j] - 1; \\
 \hphantom{140}      \textbf{sumA += value; sumB += value; sumC += value;} \\
}

This manual transformation added 3 PHI nodes to the code 
changing MILEPOST feature \emph{ft29} from 0 to 1, but without any effect on the speedup. 
We then performed similar 
transformation in a few other species that did not influence the original speedup
while changing \emph{ft29} to non-zero value and thus
invalidating original decision, i.e. showing that \emph{ft29} is not relevant
to a given optimization cluster. 

When exposing this problem to machine learning specialists, we realized that 
a high prediction accuracy can be explained by finding 
meaningless correlations in a very large and sparse feature space for just a few species
that are invalidated on a very large and diverse training set.
Though relatively naive, this example highlights the importance
of creating a common repository of software species together with
their features.
Furthermore, this example shows that negative results (mispredictions
or unexpected behavior), usually overlooked by our
community, are very important in practice and should also be 
reported, shared and published to improve machine learning techniques.
For example, our colleagues from STMicroelectronics shared their counter 
examples as new software species thus contributing to a public and realistic benchmark.

Finally, cM allows to share various machine learning models thus helping researchers 
switch from showing that \emph{it is possible to predict some optimizations
based on some features} to sharing best performing predictive models 
and classifiers, and \emph{focusing on finding why they are mispredicting
together with missing features}.
It is even possible to perform continuous competitions between shared models
to balance prediction accuracy, size and speed or reduce complexity 
depending on usage scenarios. 
For example, deep neural networks can be very good in terms of prediction accuracy, 
but are also very large, costly, and, unlike simpler decision trees, do not necessarily 
help software or hardware designers understand and solve their performance problems
(i.e. black box approach).

\section{Learning data set features to enable adaptive software}
\label{sec:features}

\begin{figure}[htb]
  \centering
  \includegraphics[width=6.0in]{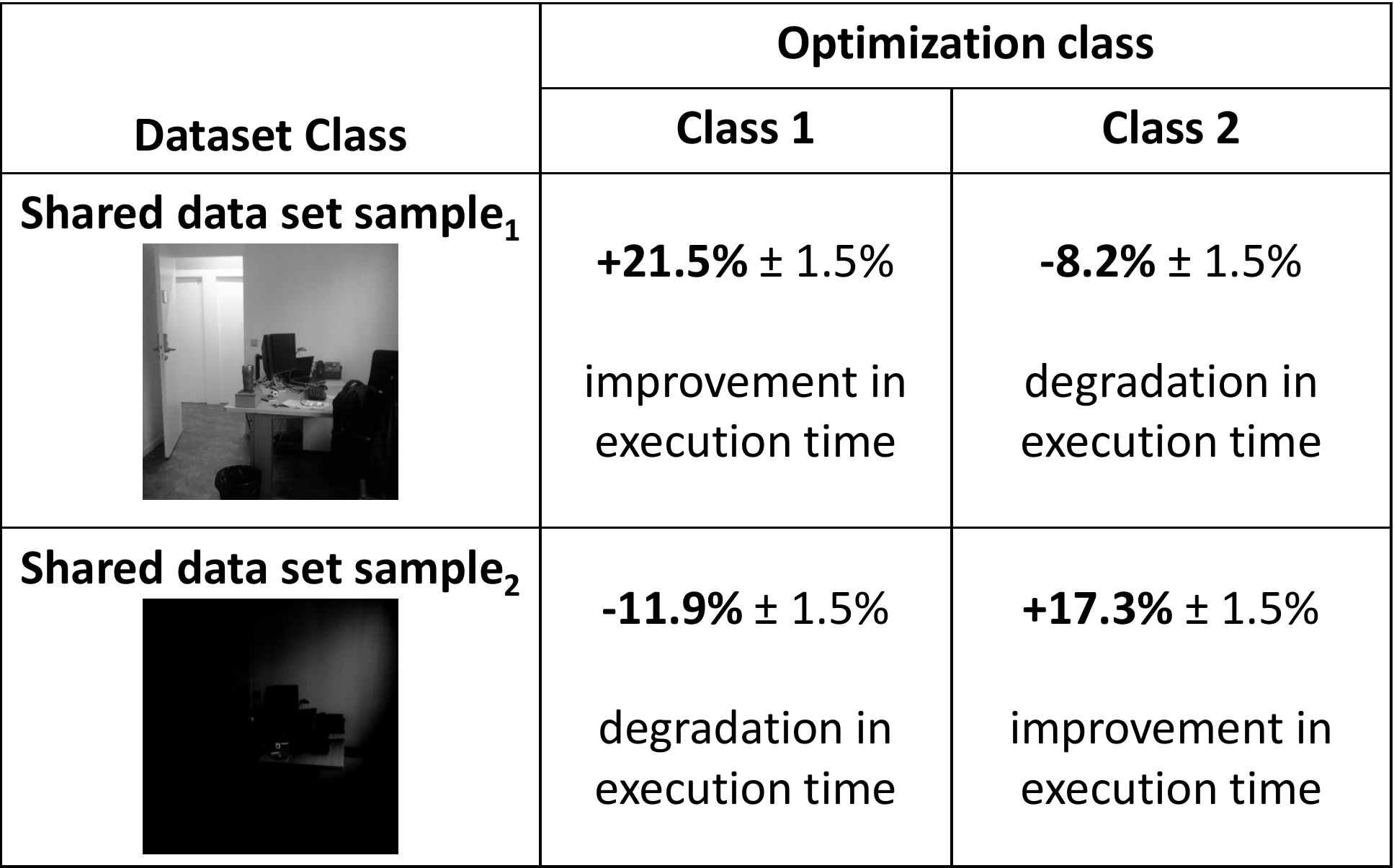}
\caption{Detecting missing data set feature "time of the day" with the help of the
community. Such feature enables adaptive software species that performs well
across all inputs.}
  \label{fig:learning-features}
\end{figure}
Though we demonstrated how our approach and methodology can help automate classification 
of shared software species to improve optimization
predictions, it still did not solve
another fundamental problem of static compilation - lack
of run-time information.
On the other hand, since cM continuously records unexpected
behavior, it helped to automatically detect that one of the real
customer's software species (image B\&W threshold filter from
a surveillance camera application similar to one shown
in Figure~\ref{fig:neural-network}) requires two distinct
optimizations with around 20\% improvement in execution
time on Intel Core i5-2540M across all shared images (data set
samples) as shown in Figure~\ref{fig:learning-features}.

In order to understand such behavior, we can now reuse the same clustering methodology
to classify available data sets and detect their features that
can explain such behavior and separate optimization classes.
Compiler designers again helped us analyze this software species and
gradually identified a suspicious "sub-species" causing unusual
behavior: \emph{(temp1~$\rangle$~T)~?~255~:~0}.
One optimization class included "if conversion" transformation
which added several predicated statements that may degrade
performance if additional branches are rarely taken due to 
a few additional useless cycles to check branch condition.
At this stage, compiler designers concluded that it is a well-known
run-time dependency which is difficult or even impossible
to solve in static compilers.
Nevertheless, one of the volunteers noticed that some images 
shown in Figure~\ref{fig:learning-features} where taken during the day 
and some during the night.
This helped us find new, simple and relevant feature related to both data set and the environment
state \emph{"time of the day"} 
that effectively separated two optimization classes. 

\begin{figure}[htb]
  \centering
   \includegraphics[width=5.6in]{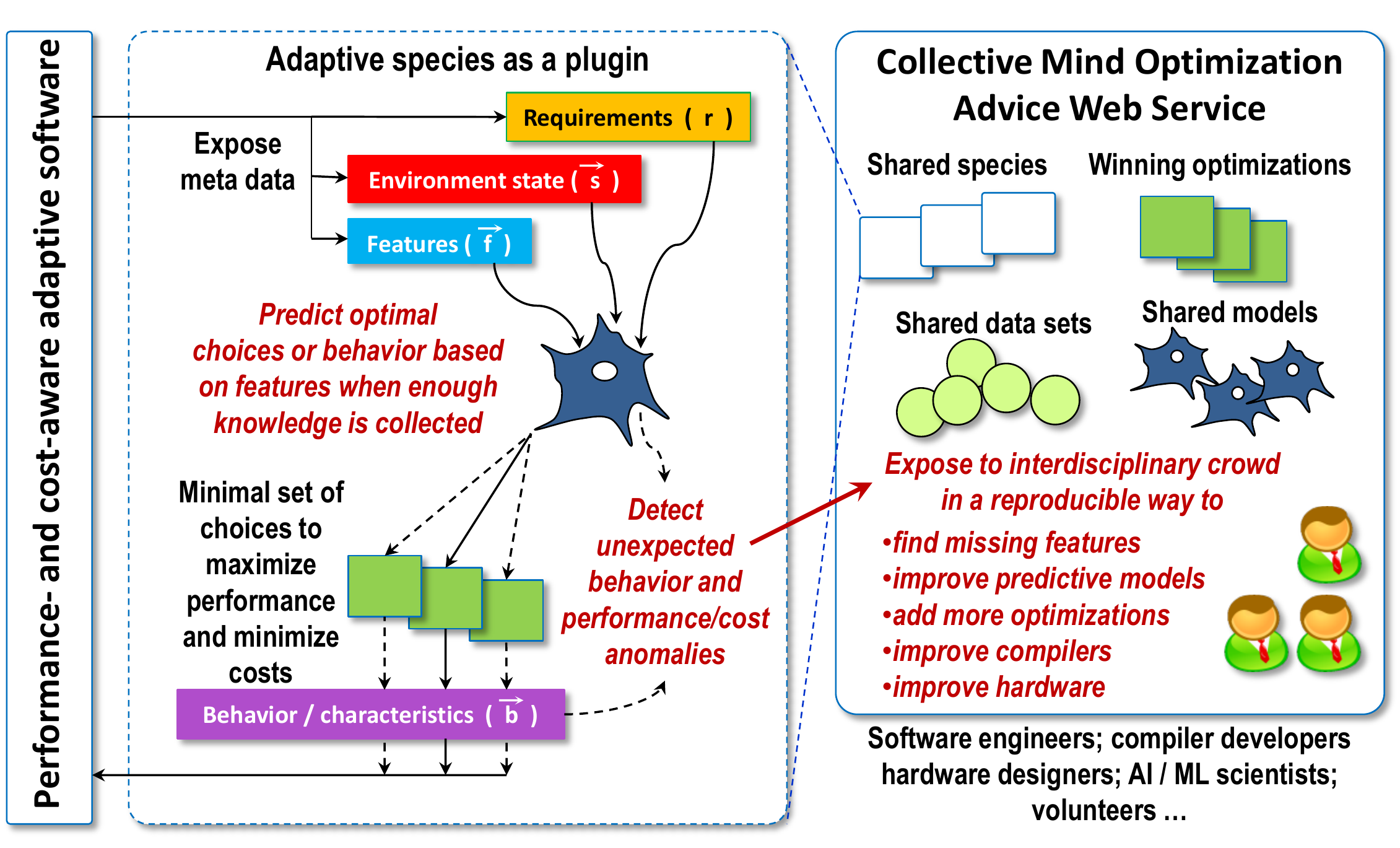}
   \caption{Concept of performance- and cost-aware self-tuning software assembled from cM plugins.
   }
  \label{fig:cm}
\end{figure}
This real example demonstrates how our
approach can help \emph{collaboratively find missing and nontrivial 
features that may not be even exist and have to be exposed} to improve
optimization prediction.
Furthermore, our approach helped substitute the threshold
filter in the customer's real software by a shared cM plugin consisting
of two differently optimized clones of this filter and a compact
decision tree.
This decision tree selects an appropriate clone at run-time based on
features of a used data set, hardware and environment state.
Therefore, our Collective Mind approach can also help make 
statically compiled software easily adaptable to different contexts 
as conceptually shown in Figure~\ref{fig:cm}.
Moreover, such software will be continuously optimized with the
help of the community while maximizing its performance, minimizing
development costs, improving productivity of software engineers
and reducing time to market.
Interestingly, cM approach can also help solve "big data problem" that 
we experienced in our first public cTuning framework~\cite{Fur2009,29db2248aba45e59:a31e374796869125}.
Rather than collecting and preserving all possible information from participating
users, we can validate incoming data against existing models and 
save only unexpected behavior. 
%

We believe that presented approach can eventually enable 
performance- and cost-aware software engineering.
We envision that instead of struggling integrating various ad-hoc optimization
heuristics to their software projects similar
to one shown in Figure~\ref{fig:my-decision-tree}, 
engineers will simply need to expose various features from data sets, software,
hardware and environment state for their software pieces.
These features will be then correlated with the winning optimizations 
either automatically or with the help of the community to gradually 
minimize execution time, power consumption, code size, compilation time, 
faults, and other costs.

\section{Conclusions and Future Work}
\label{sec:conclusion}

The computer engineering community has been desperately trying to find
some practical ways to automatically improve software performance
while reducing power consumption and other usage costs across
numerous and rapidly evolving computer systems for several
decades~\cite{xciteulike:1671417,
Dongarra:2011:IES:1943326.1943339, prace,
Hall:2009:CRN:1461928.1461946, hipeac_roadmap}.
In this paper, we presented a novel and practical approach
inspired by natural sciences and Wikipedia that may help
collaboratively solve this problem while improving productivity
of software developers.
The biggest challenge in this approach is to connect together, systematize and make practical various
techniques and tools from different interdisciplinary domains often overlooked by our community
into a coherent, extensible and top-down optimization and
classification methodology.

The backbone of our approach is a public repository
of optimization knowledge 
at \href{http://c-mind.org/repo}{c-mind.org/repo}.
It allows the software engineering community to gradually share
their most frequently used software pieces (computational species) 
together with various possible inputs and features.
All shared species are then continuously and randomly optimized
and executed with randomly selected inputs either as standalone 
pieces or within real software across numerous mobile phones, 
laptops and data centers provided by volunteers using our recent
Collective Mind framework (cM).
In contrast with a very few existing public repositories, notably
SPEC and Phoronix benchmarking platforms~\cite{spec2000,
openbenchmarking}, cM also continuously classifies best found
optimizations while exposing unexpected behavior in a reproducible way.
This, in turn, allows the interdisciplinary community to 
collaboratively correlate found classes with gradually exposed 
features from the software, hardware, data sets and environment state
either manually or using popular big data predictive 
analytics~\cite{citeulike:873540,DBLP:books/ms/4paradigm09}.
Resulting predictive models are then integrated into cM plugins
together with several pre-optimized (specialized) versions of
a given species that maximize performance and minimize costs
across as many inputs, hardware and environment states
as possible, as described in \cite{LCWP2009}

Software engineers can now assemble self-tuning applications just
like "LEGO" from the shared cM plugins with continuously
optimized species.
Such software not only can adapt to the running hardware and context,
but also continue improving its performance and minimize usage costs
when more collective knowledge is available.
This can help change current computer engineering methodology
since software engineers do not have to wait anymore until
hardware or compilers become better.
Instead, the software engineering community gradually creates a large, 
diverse and realistic benchmark together with a public and continuously 
improving optimization advice system that helps improve and validate 
future compilers and hardware.
For example, we envision that our approach will also help simplify
compilers and convert them into generic libraries of code
analysis, optimization and generation routines orchestrated
by cM-like frameworks.

To avoid the fate of many projects that vanish shortly after 
publication, we agreed with our partners to share most of the
related code and data at our public optimization repository
to continue further community-driven developments.
For example, with the help of our supporters, we already shared
around 300 software species and collected around 15000 possible
data sets.
At the same time, we also shared various features as cM meta-data
from our past research on machine learning based optimization
including MILEPOST semantic code
properties~\cite{29db2248aba45e59:a31e374796869125}, code
patterns and control flow graph extracted by our GCC/LLVM
Alchemist plugin~\cite{fursin:hal-01054763}, image and matrix
dimensions together with data set sizes from \cite{LCWP2009},
OS parameters, system descriptions, hardware performance
counters, CPU frequency and many other.

Public availability of such a repository and open source
cM infrastructure allowed us to validate our approach
in several major companies.
For example, we demonstrated how our industry colleagues managed to 
enhance their in-house benchmarking suites to considerably improve optimization 
heuristics of their production GCC compiler for a number of ARM 
and Intel based processors while detecting several architectural 
errors during validation of new hardware configurations.
Finally, presented approach helped to convert an important customer
statically compiled image processing application into
a self-tuning one that maximizes
performance to reach real time constraints and minimize all other
costs including energy, overall development and tuning effort,
and time to market.

As a part of the future work, we plan to simplify as much
as possible the experience of software engineers and volunteers
wishing to participate in our project.
Therefore, we are currently extending our cM framework
to automate identification, extraction and sharing of the
frequently used and most time consuming software pieces and their
features in real programs.
For this purpose, we plan to use and extend our Interactive
Compilation Interface for GCC and LLVM while connecting
cM framework with Eclipse IDE~\cite{eclipse} to simplify
integration of our cM wrappers and performance/cost monitoring
plugins with real applications, with Docker~\cite{docker} and
CARE~\cite{Janin:2014:CCA:2618137.2618138} to automatically
detect all software dependencies for sharing, and with Phoronix
open benchmarking infrastructure~\cite{openbenchmarking} to add
even more realistic software pieces to our repository.

At the same time, our top-down methodology originating from
physics allows the software engineering community benefit from all
existing optimizations including powerful polyhedral
source-to-source code restructuring and
parallelization~\cite{uday08pldi} by gradually adding them to our
cM performance tracking and tuning
framework~\cite{fursin:hal-01054763}.
Furthermore, researchers now have an opportunity to immediately
validate their novel or existing optimization techniques across
a realistic benchmark and a large number of participating
computer systems.
Unified "big data" repository of optimization knowledge also helped
us initiate collaboration with AI and physics departments
to gradually characterize complex interactions between shared
software pieces inside large applications using agent based
techniques~\cite{Shoham:2008:MSA:1483085}.
We are also actively working with the academic and industrial
community through the cTuning foundation to continue adding 
more coarse-grain and fine-grain species to our repository 
to gradually cover all possible and frequently used software.
For example, we would like to add algorithmic species 
from~\cite{nugteren2013algorithmic} to our system and 
continuously optimize and characterize them.
We hope, that in a longer term, our approach will help software
and hardware engineers boost innovation and focus on implementing 
new ideas and functionality rather than worrying about performance
regressions and costs.
It should also help make machine learning for compilation and architecture 
practical (including autotuning and run-time adaptation particularly
important nowadays for mobile devices and data centers) while avoiding
common pitfalls such as using a small subset of non-representative 
data sets and features.
Interestingly, our approach may also eventually help make brain inspired 
computing practical since we can continuously collect large number
of realistic data sets, continuously improve predictive models and
collaboratively find missing features (to some extent enabling
large and distributed brain).
Finally, we plan to use our public repository to promote and
support initiatives on artifact evaluation as a part
of reproducible and sustainable software 
engineering~\cite{new_pub_model}.

\section{Acknowledgments}
\label{sec:ack}

Presented work was supported by the HiPEAC, STMicroelectronics,
cTuning foundation, EU FP7 609491 TETRACOM project and ARM.
We would like to thank Ed Plowman (ARM), Marco Cornero (ARM) 
and Sergey Yakushkin (Synopsys) for interesting feedback 
and discussions. 



\bibliographystyle{abbrv}
\bibliography{article}

\newpage
\section{Appendix:
Collective Knowledge -- a customizable knowledge management framework for systematic and collaborative experimentation}
\label{sec:ck}

\begin{figure}[htb]
  \centering
   \includegraphics[width=5.0in]{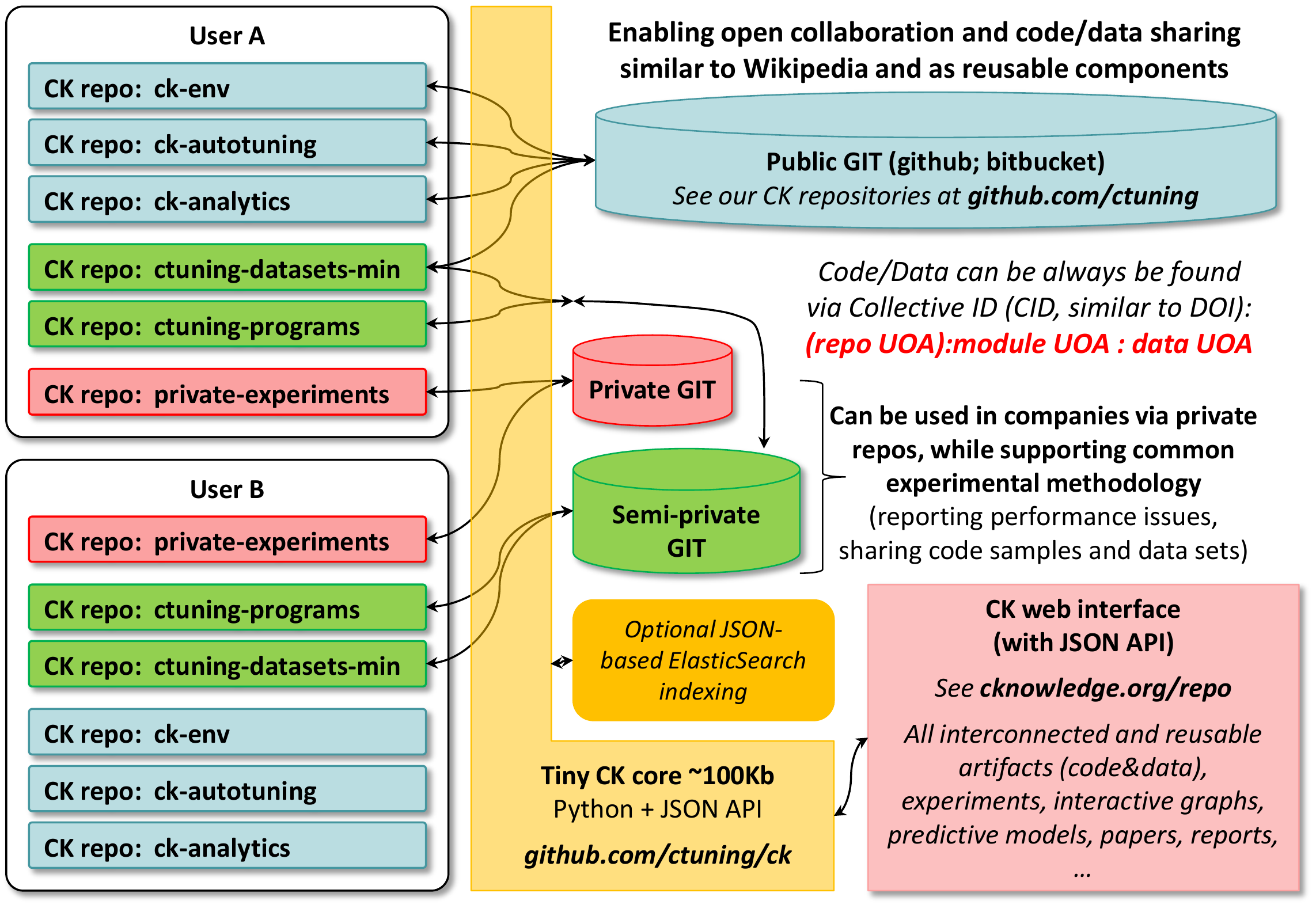}
  \caption{
   Collective Knowledge (CK) -- a new framework for systematic and collaborative experimentation.
  }
  \label{fig:ck}
\end{figure}

Over the past two years, we have collected considerable feedback from
academic and industrial users about Collective Mind, the third
version of the cTuning technology. 

Positive feedback highlighted considerable improvements over the previous
versions of the cTuning technology~\cite{Fur2009} for sharing experimental
artifacts such as programs, data sets, graphs and models within work groups
(e.g. at the same institution) and with a wider community (e.g. alongside
publications).
Users particularly appreciated the flexibility and ease of organizing their
knowledge using an open JSON-based repository format and API rather than being
forced to use rigid and possibly closed database schemas.
Users were also intrigued by the possibilities for crowdsourcing exploration of
large multi-dimensional optimization spaces and applying machine-learning
techniques to generate and reuse knowledge across many platforms (phones,
tablets, laptops, cloud services and supercomputers).

On the other hand, users disliked having to download over 20 MB to obtain the
framework with all components, having to use the web service instead of plain
command line, slow invocation and slow processing when ElasticSearch indexing
was unavailable.

After carefully considering this feedback and available options, we decided to
develop a new version from scratch while keeping the original open JSON-based
repository format and API.
Thanks to the 6-months grant from the EU FP7 609491 TETRACOM project, we have
developed Collective Knowledge (or CK for short), the fourth version of the
cTuning technology, publicly available at \url{http://github.com/ctuning/ck}
under permissive BSD license.
The main focus was to make the core of Collective Knowledge as small, fast,
simple and command-line-oriented as possible, while moving all additional
functionality such as predictive analytics, autotuning, statistical analysis
and web services into separate CK plugins shared via GitHub, Bitbucket or
similar public services as conceptually shown in Figure~\ref{fig:ck}.

The Collective Knowledge core works about 10 times faster than Collective Mind,
is only around 100 KB in size, has minimal dependencies (mainly Python and
Git), can be installed as a standard Python package and can be easily
integrated with IPython Notebook.
Furthermore, users can download only needed packages (while automatically
resolving dependencies on other repositories) using only one shell command
\emph{ck pull repo:[repo alias] (--url=[Git url])}.
It is also possible to update the whole framework and all shared repositories
at any time using \emph{ck pull all}.

We believe this has considerably simplified using the cTuning technology and
methodology for collaborative projects, including sharing of experimental
artifacts, interactive reports and publications. (Examples are available at
\url{http://cknowledge.org/repo}.)

\begin{figure}[htb]
  \centering
   \includegraphics[width=7.0in]{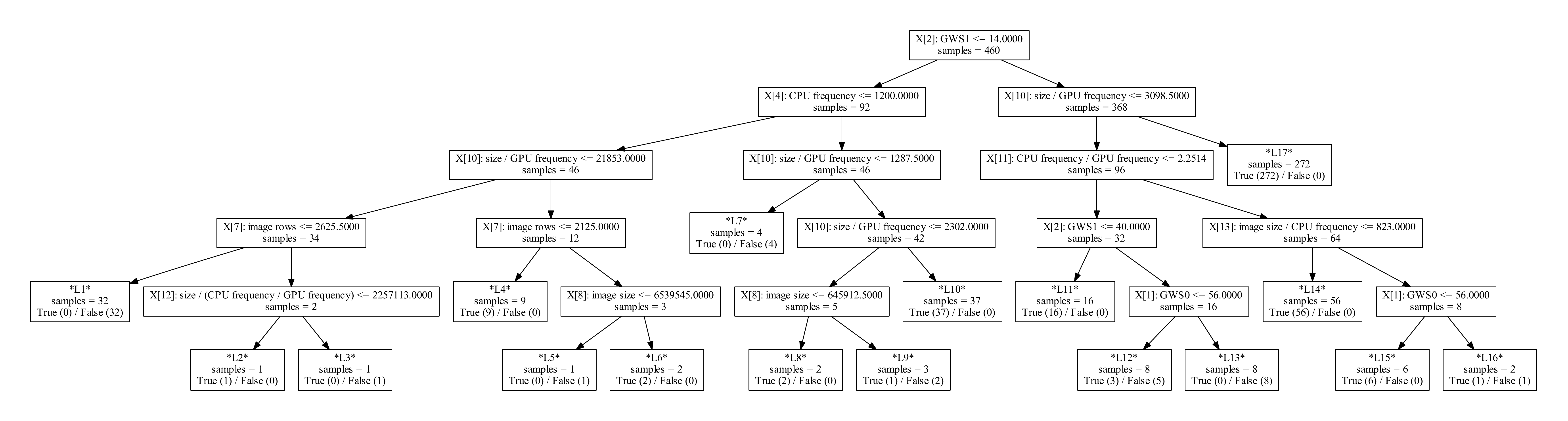}
  \caption{
    Example of a decision tree built online for predictive
    scheduling of an OpenCL-based video processing kernel (active
    learning) on Samsung Chromebook 2 with MALI GPU and ARM-based
    CPU depending on data set features, CPU/GPU frequency and
    other parameters.
  }
  \label{fig:ck_decision_tree}
\end{figure}

For example, as a part of our long-term and continuing effort to share all
code, data, models and interactive graphs from our research in a reproducible
way~\cite{Fur2009}, we are gradually converting all our past experiments and
related artifacts to the new CK format.
Our current focus is to use CK to solve old but still acute problems of
hardware benchmarking, program autotuning, and run-time adaptation using
collective knowledge and predictive models.
Some artifacts from this paper are already available in the shared CK
repository which can be obtained via \emph{ck pull repo:reproduce-ck-paper}.
For example, we have reproduced our past work on adaptive CPU/GPU
scheduling~\cite{JGVP2009} for an OpenCL-based video processing kernel on the
Samsung Chromebook 2 powered by the ARM Cortex-A15 CPU and
ARM Mali-T628 MP6 GPU.
This includes continuous and online exploration and tuning of OpenCL parameters
such as the local work size versus the computational intensity, video frame
features, and processor frequencies.
At the same time, we have applied active learning as in~\cite{JGVP2009} to build and
gradually refine a decision tree (shown in Figure~\ref{fig:ck_decision_tree}
and shared at
\url{http://cknowledge.org/repo/web.php?wcid=84e27ad9dd12e734:f4b5b8fded3ce933})
that could effectively predict whether to schedule the main kernel on the GPU
(true) or the CPU (false) to minimize the overall execution time (improve the
number of frames per second).

We believe that CK will improve support for our long-term vision towards
collaborative, systematic, and reproducible computer engineering combined with
statistical (``machine learning'') techniques from
physics~\cite{Fur2009,fursin:hal-01054763} and powerful predictive analytics
frameworks such as SciPy, R, IBM Watson, Google Brain or similar.
Furthermore, we envision that our approach will allow engineers to focus on
quick prototyping of ideas, knowledge discovery and innovation, while helping
them to get right complex, time-consuming and error-prone issues of
experimental design. 

To some extent our approach is similar to Wikipedia allowing users to share and
continuously improve knowledge and experience while collaboratively extending
meta-data, stabilizing API, fixing errors, improving predictive models, finding
missing features and many more.
In fact, all CK entries can be referenced by a unique collective identifier
CID similar to DOI (repository UID or alias:module UID or alias:data UID or
alias), and already include a dedicated wiki page (thus allowing researchers
to collect feedback about their shared artifacts and engage in discussions
with the community).

Finally, we promote CK to enable Artifact Evaluation initiatives at leading
academic conferences and community-driven reviewing~\cite{new_pub_model},
along with other useful tools such as Docker.

Further information about CK is available at \url{http://github.com/ctuning/ck/wiki}
and \url{http://github.com/ctuning/ck}.

\end{document}